\documentclass{aa}
\usepackage{amsmath,graphicx,natbib,tabularx}

\bibpunct{(}{)}{;}{a}{}{,} % Astronomy & Astrophysics, ApJ

\newcommand{\aba}{AB Aurig\ae}

\begin{document}  

\title{An emission ring at 20.5 $\mu$m around the HAEBE star\\  \aba: unveiling the
       disc structure} 

\titlerunning{An emission ring around \aba}
\author{E.\,Pantin\inst{1} \and J.\,Bouwman\inst{1} \and P.O\,Lagage\inst{1}} 
\institute{DSM/DAPNIA/Service d'Astrophysique, CEA/Saclay, F-91191
 Gif-sur-Yvette, France}
\offprints{E. Pantin} 
\mail{epantin@cea.fr}
\date{Submitted to Astronomy \& Astrophysics Letters}

\abstract{
Isolated HAEBE stars are believed to represent an intermediate stage
of objects between young stellar objects surrounded by massive,
optically thick, gaseous and dusty disks and Vega like stars
surrounded by debris disks. The star \aba \ is already known for being
surrounded by an intermediate-stage dust disk emitting a fairly large
infrared and (sub-)millimetric excess. Until now, the outer disk
structure has only been resolved at millimeter wavelengths and at
optical wavelength coronographic imaging. We have obtained 20~$\mu$m
images which show an unexpected ellipse-shaped disk structure in
emission at a distance of about 260 AU from the central star.  Large
azimuthal asymmetries in brightness can be noticed and the center of
the ellipse does not coincide with the star.  A simple, pure
geometrical model based on an emission ring of uniform surface
brightness, but having an intrinsic eccentricity succeeds in fitting
the observations.  These observations give for the first time clues on
a very peculiar structure of pre-main-sequence disk geometry, i.e. a
non uniform increase in the disk thickness unlike the common usual
sketch of a disk with a constant flaring angle.  They provide also
valuable informations on the disk inclination as well as its dust
composition; at such a large distance from the star, only transient
heating of very small particles can explain such a bright ring of
emission at mid-infrared wavelengths.  Finally, the increase of
thickness inferred by the model could be caused by disk instabilities;
the intrinsic eccentricity of the structure might be a clue to the
presence of a massive body undetected yet.

\keywords{Circumstellar matter -- Stars: formation -- Stars: pre-main-sequence}}

\maketitle

\section{Introduction}

%The discovery of infrared excesses around a noticeable fraction of
%main-sequence stars \citep[e.g.][]{aumann1984,Gillett1986,habing2001}
%and attributed to the presence of cool dust grains orbiting the star
%and geometrically arranged in a disk, has triggered an intense study
%of these objects since these disks may be linked to planetary formation (see e.g. 
%\citet{2000prpl.conf..639G}). 

%Recent discoveries of ``young
%debris'' or ``old'' YSO -- the question is still debated --
%circumstellar disks around relatively ``old'' and isolated (i.e. not
%associated to any star forming region) stars have shown that this
%phenomenon extends towards Pre-Main-Sequence (PMS) stars (see the
%photometric survey by \citet{malfait1998}, see also the study of
%spectral characteristics by \citet{HerbigOverview}). 

%These disks are
%in an intermediate stage between dense, gas rich, optically thick
%disks observed around young stellar objects (YSO) (aged of say less
%than 1 million years), and Vega like disks. Although the
%starlight is usually absorbed by less than a few magnitudes, even
%sometimes not attenuated at all when the disks are observed face-on,
%their inner disks contain still relatively dense, optically thick
%material, while outer parts are believed to be optically thin.

Herbig Ae/Be (HAEBE) stars represent a class of intermediate mass, pre-main-sequence (PMS) stars,
first described as a group by \cite{herbig1960}.
The circumstellar (CS) disks found around these stars are
believed to be the sites of on-going planet formation. 
%As these disks
%evolve with time, the sub-micron sized dust grains present at the
%formation time of the disks coagulate to form larger objects and
%eventually planet(esimal)s. 
By studying the characteristics and evolution of the CS disk
and its dust composition, valuable insights can be obtained into the
processes leading to the formation of planets, and put
constraints on disk and planet formation models. 
Infrared spectroscopy obtained with the Infrared Space Observatory
(ISO) has given us insight into the dust composition of sample of isolated HAEBE systems
\citep[e.g.][]{abaur,processing,HerbigOverview}. While these spectra reveal a rich mineralogy, no direct
information concerning the spatial distribution of the different dust
species can be inferred from the ISO data.
Most studies so far have used the available spectral energy distributions (SEDs) to
put constraints on the spatial distribution of the CS material.
Models for passively heated disks surrounding PMS stars are successful in reproducing the ISO
spectra \citep[e.g.][]{dullemond2001, dominik2003},
but these models cannot be uniquely constrained from SED fitting alone \cite{degenaracy}, for this 
spatially resolved imaging, as presented in this paper, is required.
\begin{figure*}[t]
\vspace*{-3cm}
\parbox{11.0cm}{
\begin{center}
\vspace*{2.5cm}
\resizebox{13.0cm}{!}{\includegraphics[angle=90]{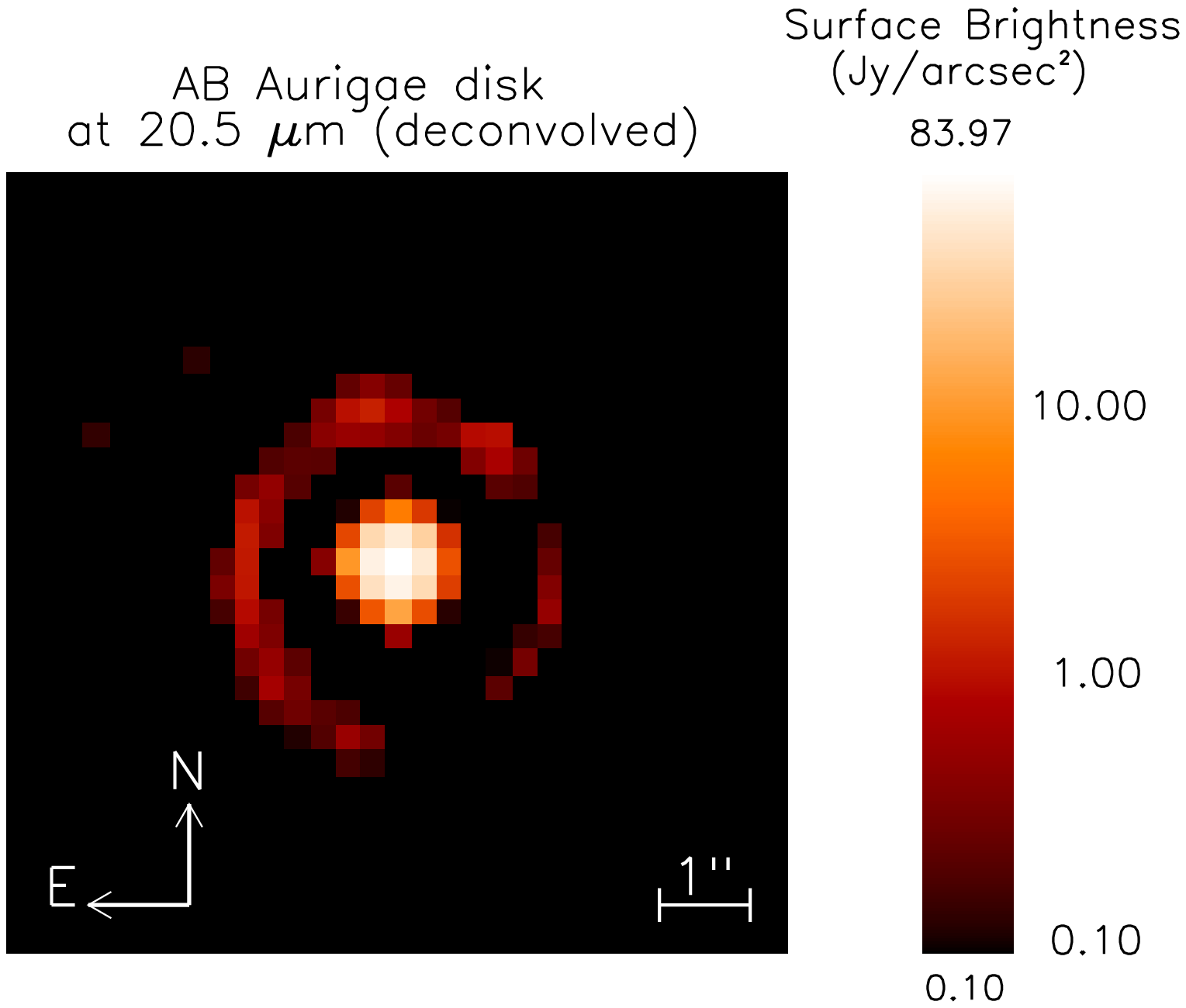}}
\end{center}
}
\hspace*{-0.5cm}
\parbox{6.5cm}{
\begin{center}
\vspace*{2.5cm}
\resizebox{7cm}{10cm}{\includegraphics[bb= 5cm 10.3cm 16cm 29.2cm,clip]{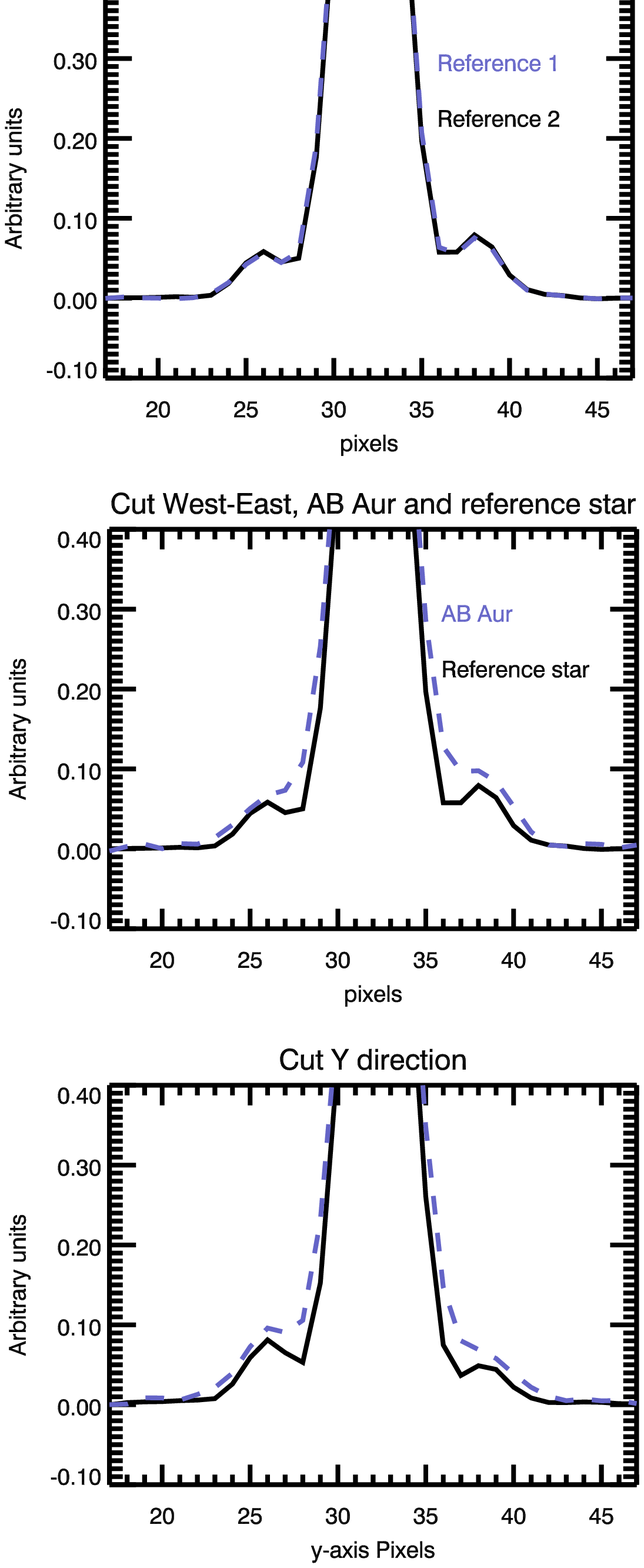}}
%\vspace*{-1.cm-}
%\resizebox{7cm}{6cm}{\includegraphics[angle=90, bb= 12cm 13cm 23.5cm 22.2cm,clip]{oplot_reference_abaur.ps}}
%\vspace*{-1.cm}
\end{center}
}
%\vspace*{-0.5cm}

%\begin{center}
%\setlength{\unitlength}{1cm}
%\begin{picture}(16,13)(0,0)
% \put(0,0){\resizebox{130mm}{!}{\includegraphics[angle=90]{image.ps}}}
% \put(8,0){\resizebox{75mm}{!}{\includegraphics[angle=90, bb= 12cm 11.5cm 24cm 22cm,clip]{oplot_references.ps}}}
% \put(8,6.5){\resizebox{75mm}{!}{\includegraphics[angle=90, bb= 12cm 11.5cm 24cm 22cm,clip]{oplot_reference_abaur.ps}}}
%\end{picture}
%\end{center}
\label{fig-image}
\caption{Image of the AB Aurigae disk at 20.5~$\mu$m. The left figure shows
          the deconvolved image, with a pixel scale of
          0.3\arcsec/pixel.  Clearly visible is a resolved central
          emission region surrounded by a ring like structure.  The
          panels on the right show the normalized intensity profiles
          along a cut through the CAMIRAS images of 2 reference stars
          and AB~Aur. The upper right panel shows the comparison
          between two observations of reference stars, demonstrating
          the stability of the PSF. The panel on the lower right shows
          a comparison of AB~Aur with the PSF, clearly showing that
          the central emission is extended and that a ring-like
          emission structure is also detected in non deconvolved data.  }
\end{figure*}

Among isolated HAEBE systems, the disk around \aba \ is
one of the most interesting and studied. Its star
has a probable age of 2~Myr \citep{mario1997}, indicating that,
according to current planet formation theories, planet building could
still being ongoing in this system.  ISO spectra show strong PAH
emission bands, and emission from silicates, and carbonaceous dust
grains \citep{mario_abaur,abaur}.  Though the inferred grain sizes are
differing from interstellar grains, the dust around AB~Aur seems
to be relatively unprocessed, indicating an evolutionary young system.
This seems also to be confirmed by combined ISO-SWS and sub-millimeter
observations of H$_2$ and CO rotational lines, demonstrating that the
disk still has a large gas content \citep{Thi2001}.  Its disk has been
resolved in the millimeter range by \citet{mannings1997}, showing a
structure consistent with a Keplerian disk.  Its surrounding nebulae
and outer disk structure were also studied in the visible range using
broad-band coronographic observations \citep{grady_stis_abaur},
showing a disk with spiral-shaped structures. Near-IR interferometric
\citep{millan1999, millan2001, eisner2003}, have resolved the inner
parts of the disk, showing it to be consistent with a passive disk
with an inner hole, seen at a low ($\le 40\degr$) inclination angle. 
Recent mid-IR imaging
\citep{chen2003}, has also resolved the inner structure, showing it to
originate from thermal emission from dust grains heated by the stellar
radiation field near the central star. Here we present mid-IR
imaging, not only resolving the thermal emission from dust close to
the central star, but also an emission structure at the outer parts of
the disk. In the next sections we will describe these observations
and propose a pure geometrical model for explaining our results, unlike
the paradigm according to which gravitational interactions with
massive bodies are inferred when observing brightness asymmetries in
dusty disks.
 
\section{Data collection and Reduction} 

The observations were performed using the CEA mid-IR camera CAMIRAS 
\citep{CAMIRAS},
equipped with a Boeing 128x128 pixels BIB detector sensitive
up to a wavelength of  $\approx28$ $\mu$m.  \aba \ was observed from 
the CFH 3.6m telescope on 1999 August 1 and
between 2000 March 16 and March 20. During these two runs, seeing and
weather conditions -- humidity and amount of atmospheric precipitable
water -- were particularly favorable and
extremely stable in time.  We spent a total integration time on \aba \
of 2h split into several nights. Each dataset was reduced
independently in order to avoid any erroneous conclusion due for
instance to some corrupted dataset; it permits also to evaluate error
bars on our results. The orientation of the array on the sky was
carefully determined at the start of each observing run.  The pixel
size was $0.29''$. In the 20 $\mu$m window, we used a filter centered
at 20.5 $\mu$m and with a bandpass (FWHM) $\Delta\lambda$ = 1.11
$\mu$m. This filter is free of any important atmospheric line
contribution.
\begin{figure*}[t]
%\begin{center}
%\parbox[b]{7.5cm}{
%\resizebox{75mm}{!}{\includegraphics[angle=90]{ring_radius.ps}}
%\vspace{1.2cm}
%\resizebox{75mm}{!}{\includegraphics[angle=90]{ring_intensity.ps}}
%\vspace{0.05cm}
%}
%\parbox[b]{10cm}{ 
%\resizebox{75mm}{!}{\includegraphics[angle=90]{ring_fwhm.ps}}
%\resizebox{10cm}{!}{\includegraphics[angle=90,bb=0 0 400 800]{simulated_image.ps}}
%\resizebox{10cm}{!}{\includegraphics[angle=90]{simulated_image.ps}}
%}
%\end{center}
\vspace*{-0.75cm}
\begin{center}
\setlength{\unitlength}{1cm}
\begin{picture}(16,13)(0,0)
 \put(0,6.5){\resizebox{75mm}{!}{\includegraphics[angle=90]{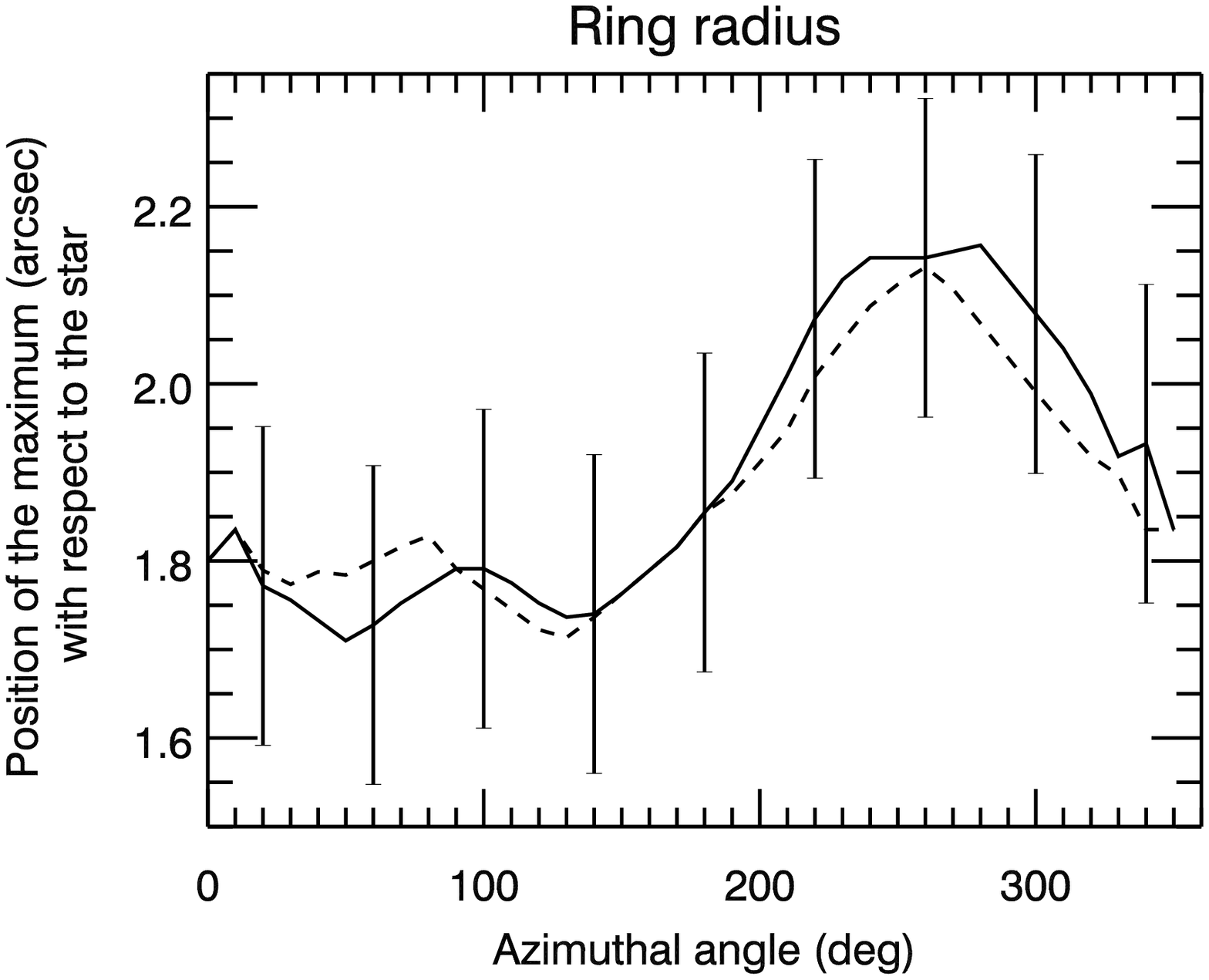}}}
 \put(0,0){\resizebox{75mm}{!}{\includegraphics[angle=90]{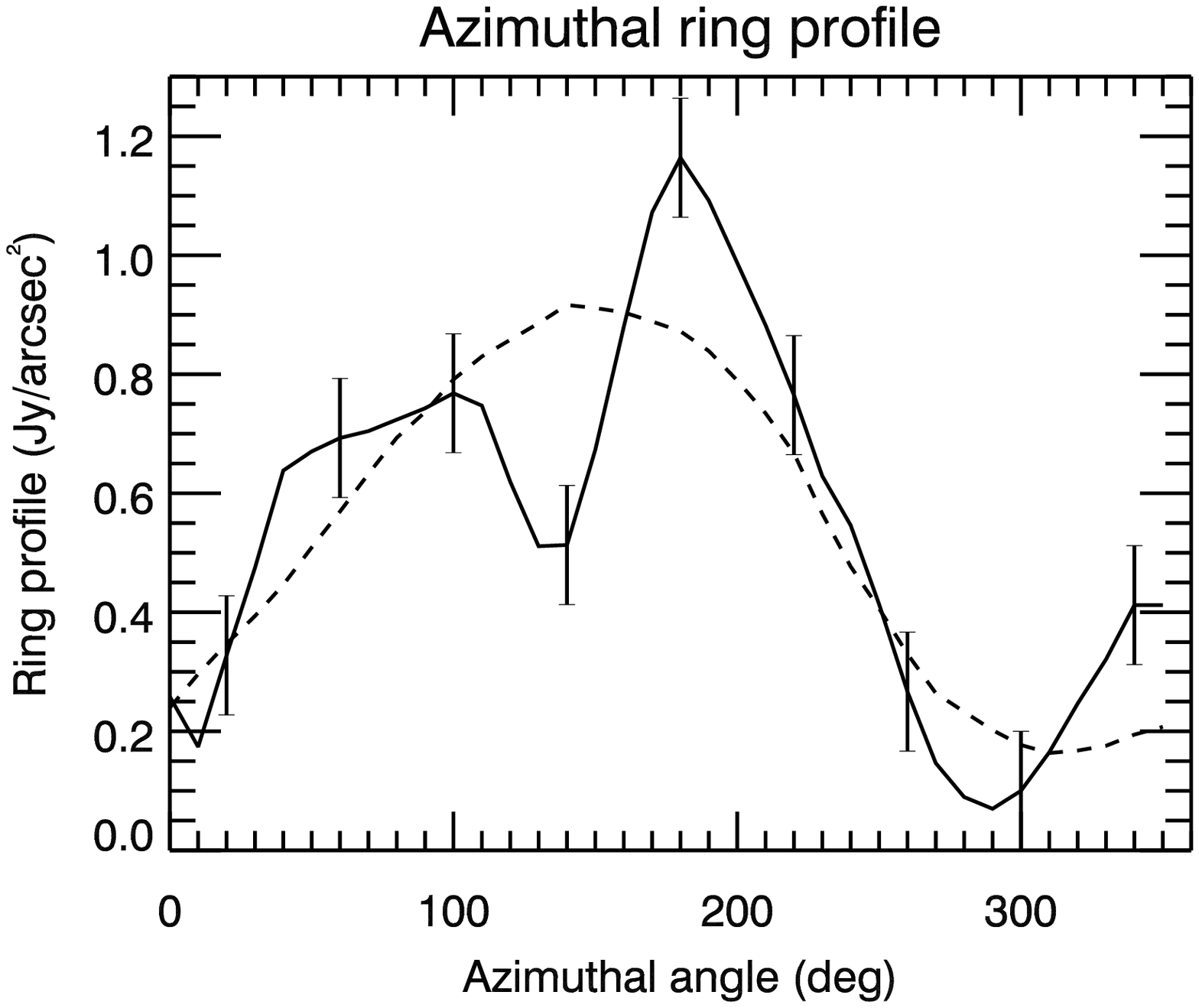}}}
 \put(8,6.5){\resizebox{75mm}{!}{\includegraphics[angle=90]{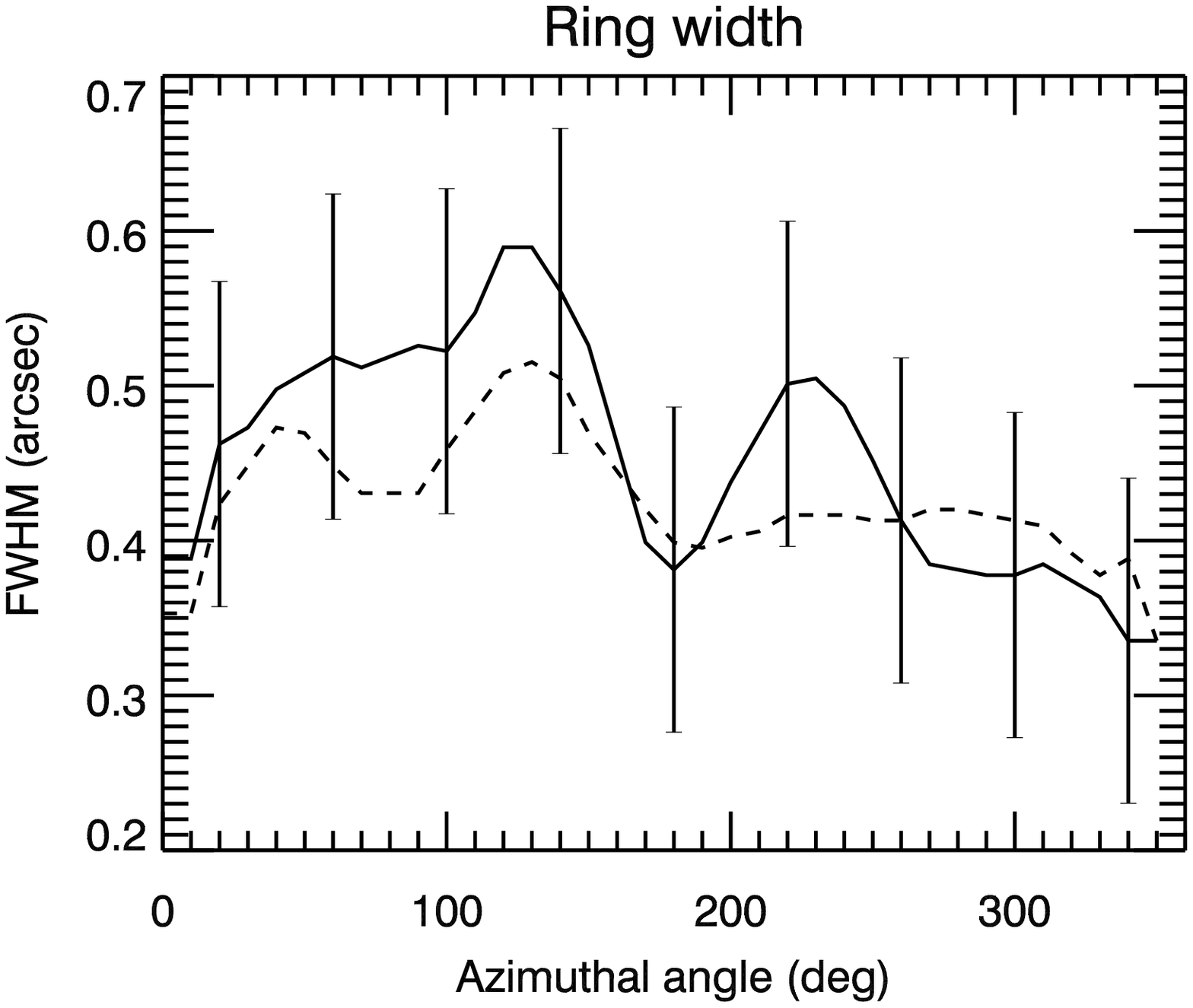}}}
 \put(8.5,0){\resizebox{10cm}{!}{\includegraphics[angle=90]{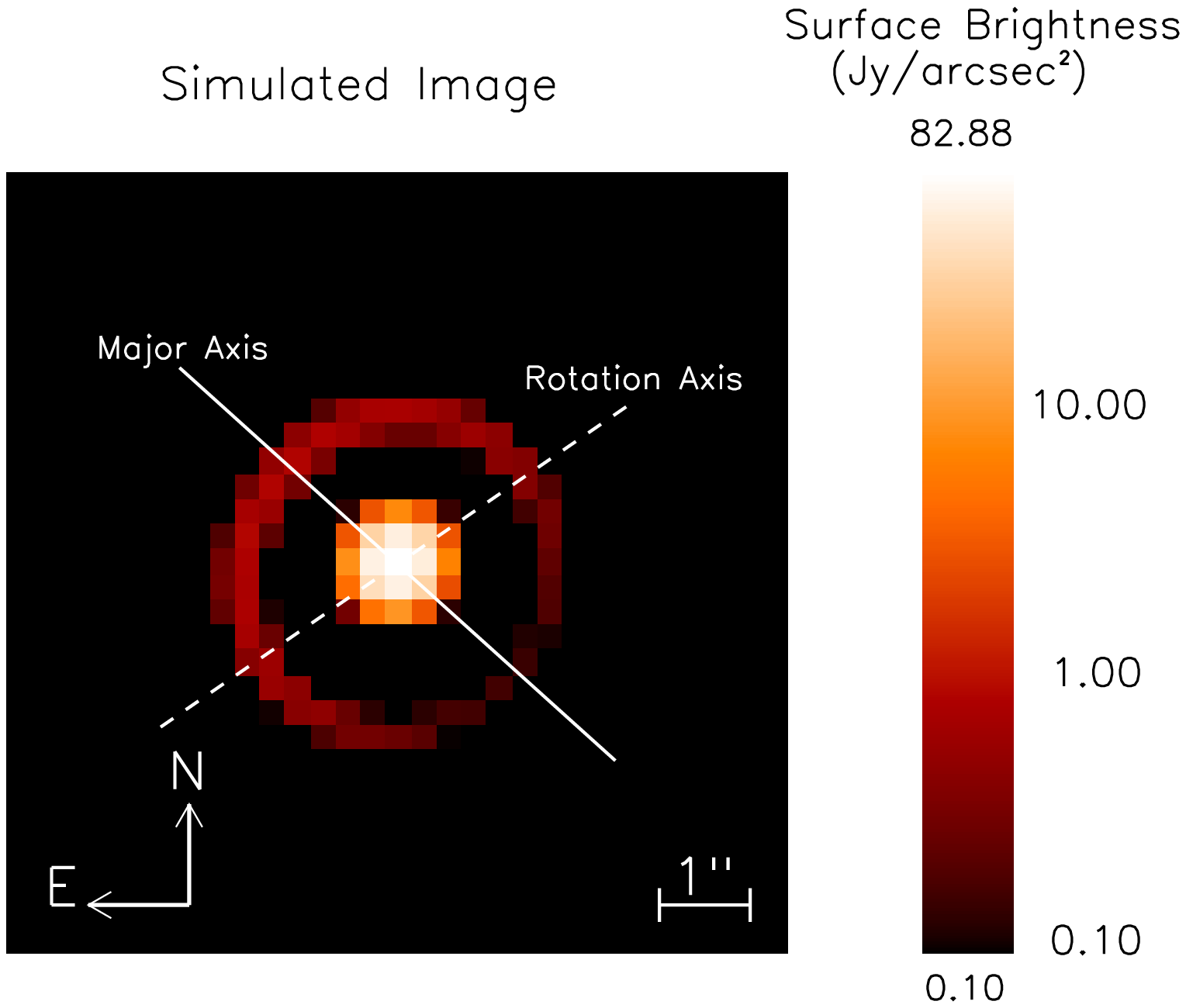}}}
\end{picture}
\end{center}
\caption{ The structure of the emission ring as an function of azimuth, compared with our 
          best-fit model. Plotted are, from left to right, top to
          bottom, with the solid lines, the ring radius, its width
          (fwhm), and its intensity profile as a function of azimuthal
          angle (from the west direction, going counter clock
          wise). Over-plotted with dashed lines are the modeled ring
          radius, width and intensity profile.  For the three solid line
          plots (data), the profiles have been smoothed by a 5 points (number
          of angle points being 36) boxcar in order to smooth high
          frequencies variations. The error bars indicate the formal
          1~$\sigma$ error.  Also shown is the best fit model image
          (bottom right) with overlaid the rotation axis (link to
           the disc inclination) and the major axis of the intrinsic
           ellipse described in section~\ref{sec:results}.}
\label{fig-profile}
\end{figure*}
%  To avoid saturation of the detector by the high ambient photon
%  background, the image elementary integration time was chosen to be
%  17.9 ms. 
  The source was always observed at an airmass of less than 1.3.
  Standard chopping and nodding techniques were
  applied with a chopping throw of 16\arcsec \ North and nodding amplitude of
  20\arcsec West. A shift-and-add procedure was applied to each final cube of
  images using a correlation based method with a re-sampling factor of
  4:1.  In order to get the best spatial resolution in the dataset,
  the nodding direction was perpendicular to the chopping direction,
  resulting in an image containing four times the image of the
  source. The four sources were extracted and co-added using a
  cross-correlation method.
  Finally, each independent dataset was deconvolved separately
  before co-addition to produce the final image.
  The peak signal-to-noise ratio in the various datasets 
  ranges from 225 to 255.
The standard star $\alpha$ Tau was frequently monitored for further
data photometric calibration and for PSF measurements.  Because final
co-added images are strongly limited in spatial resolution by the
seeing and above all by the 3.6m telescope diffraction-pattern (FWHM
of $1.4''$ at 20.5 $\mu$m), the use of deconvolution techniques is
mandatory in order to recover the best spatial resolution in MIR
images. We used the Multiscale Maximum Entropy Method developed by
\cite{Pantin1996} based upon wavelet analysis based on 
the concept of multiscale information.  Shannon theorem prescription
allows to pursue the deconvolution down to twice the pixel size, in
this case down to $\approx \;0.6''$, and iterations are stopped
according to the residual map (i.e. the data minus the re-convolved
solution), which properties should be consistent with noise
characteristics.  Each epoch dataset was processed (including image
deconvolution) independently in order to check the consistency of the
results.

\section{Results and discussion}
\label{sec:results}
\subsection{Emission structures and photometry}

After deconvolution, the 20.5~$\mu$m image of \aba \ shows two main
structures, as can be seen in the left panel of Fig.~\ref{fig-image}.
The most striking one is a narrow ring-like structure slightly open
(i.e. the flux is decreased by an order of magnitude) in the SW
direction.  It has a total flux of 2.7$\pm$0.3 Jy (taking into account
deconvolution errors) and is located at an
average distance of 1.8 \arcsec (260 AU, assuming a distance of
144~pc) and with a typical width FWHM of 0.45 \arcsec (65 AU).  %TO BE CHECKED !!  
The second structure, and brightest, with a total flux of
48.7$\pm$5 Jy, is concentrated in the close vicinity of the
star. 
%Comparing \aba \ and PSF profiles, as can be seen in the right
%panels of Fig.~\ref{fig-image}, one can clearly notice that the
By comparing \aba \ and PSF profiles we find that the 
central emission is also resolved with a FWHM (diameter) of 3 pixels
or 0.9 \arcsec (130 AU), probably due to thermally emitting dust
grains in the inner regions of the circumstellar disk (note that the 
star emission ($\approx$ 4 mJy) is negligible at that wavelength).

\begin{table}[t]
\caption[]{Best fit model parameters of \object{AB Aur}. Listed are the model parameters
defining the system orientation, the ring emission and central
emission region.}
\label{table-1}
\begin{center}

\begin{tabular}{@{} llll @{}}
\hline\noalign{\smallskip}
\hline\noalign{\bigskip}
  Parameter     &Value &$\pm$& 1$\sigma$ error  \\
\noalign{\smallskip}
\hline\noalign{\smallskip}
\multicolumn{4}{l}{system orientation$\dag$:} \\
 $\Theta$         &   65 \degr  &$\pm$& 2.0 \degr  \\
 $i$              &   -9 \degr   &$\pm$& 1.0  \degr  \\
\multicolumn{4}{l}{ring parameters:} \\
 $e$ & 0.13 &$\pm$& 0.01 \\ 
$a$ & 1.95 \arcsec &$\pm$& 0.01\arcsec \\
$\Phi$ & 129 \degr &$\pm$& 5.0 \degr \\ 
$\Delta$R & 0.12 \arcsec &$\pm$& 0.0034 \arcsec \\
 H$_0$ & 0.57 \arcsec &$\pm$& 0.038 \arcsec\\ 
$\Delta$H & 0.163 \arcsec &$\pm$& 0.012 \arcsec \\
 I$_\mathrm{ring}$& 1.98 $\mathrm{Jy}$ $ \mathrm{arcsec}^{-2}$&$\pm$&0.017 $\mathrm{Jy}$ $ \mathrm{arcsec}^{-2}$ \\
\multicolumn{4}{l}{parameters central emission:} \\
R$_\mathrm{center}$& 0.42\arcsec &$\pm$& 0.0044 \arcsec \\
I$_\mathrm{center}$& 38 $\mathrm{Jy}$ $ \mathrm{arcsec}^{-2}$&$\pm$& 1.4 $\mathrm{Jy}$ $ \mathrm{arcsec}^{-2}$ \\
\noalign{\smallskip}
\hline
\multicolumn{4}{l}{\parbox{0.9\linewidth}{\small{$\dag$angles are given in degrees counter clockwise with
respect to the west direction (right direction)}}}\\
\end{tabular}
\end{center}
\end{table}

\subsection{Interpretation: disk structure and the nature of the emitting grain population.}
The inner resolved structure, in our schematic view of the disk
thermal emission, corresponds to thermal emission of hot grains at the
inner parts of the disk, between 10 AU to about 65 AU from the star,
where the temperature of the dust grains is such that they thermally
emit efficiently at around 20~$\mu$m. We find a similar structure as
reported by \citet{chen2003}, which can be explained by thermal
emission from either a disk or an envelope.  We will discuss and model
this emission extensively in a forthcoming paper.

Here we will concentrate on the intriguing ring-like structure at
about 260 AU from the star.  First of all, ``standard'' thermal
emission from ``big grains'' is discarded at such a large distance from
the star, because the stellar radiation would heat them to a
temperature around 100~K or less.  These grains would then produce a
flux in the range 30 to 60~$\mu$m inconsistent with the spectrum
measured by ISO/SWS.
Therefore, we must infer transient heating of very small (like PAH
particles for instance) by UV radiation, that relax through emission
of narrows bands. PAH bands were already found in the SWS spectrum of
\aba \ \citep{mario_abaur,abaur}. Indeed, according to
\citet{schutte1993} and \citet{draine2001}, large PAH molecules (or
small grains), consisting of a few thousand carbon atoms, can
efficiently emit infrared radiation with a ``plateau''-like spectral
shape around 20~$\mu$m producing then the observed emission of
$\approx$ 3~Jy at 260 AU from the star; derived relative band strength
are also consistent with PAH emission models cited above.

The fact that the geometrical emission is arranged in a ring puts
strong constraints on the disk geometry.  
In the following we call apparent or projected ellipse the ellipse
directly seen on the 20~$\mu$m image.  The most constraining feature
of this ring-like elliptical structure is that it is
\emph{off-centered along the major axis of the projected
ellipse}.  In principle, inclining a circular structure leads to
a projected elliptical structure that can be off-centered but always
\emph{along the apparent minor axis}. 
Also the systematic brightness variation along the ring is not symmetric with
respect to the minor axis of the apparent ellipse meaning that either
this brightness variation is intrinsic to the ring or that the
orientation of the inclination axis is not the same as the major axis
of the ellipse.  This leads unavoidable to the conclusion that
we are looking at an intrinsically elliptical structure seen under a
small inclination angle.

\begin{figure}
\centerline{\resizebox{\hsize}{!}{\includegraphics[angle=0]{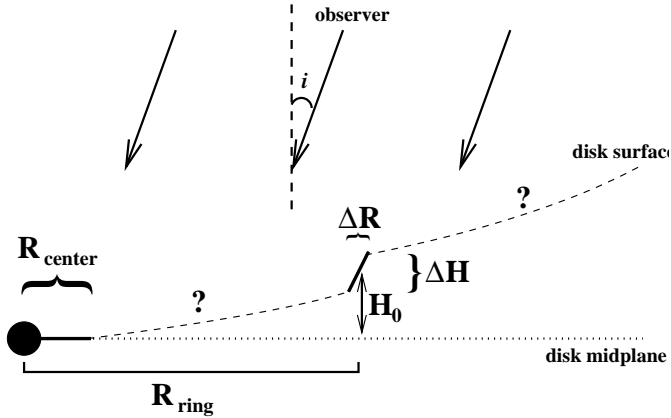}}}
\caption{Schematic representation of the disk structure in the AB~Aur system. 
Drawn is the derived geometry of the disk surface. Indicated in the
figure are the model parameters defining this geometry as listed in
Table~\ref{table-1}. R$_{\mathrm{ring}}$ is the distance to the star and varies between
$a(1-e)$ and $a(1+e)$.}
\label{fig-disk}
\end{figure}

To determine the main characteristics of the ring structure as seen in
the CAMIRAS image of \aba, we constructed a simple, geometrical
model based on the deconvolved image. 
Since the inferred very small grains need stellar UV or visible
photons to be excited, their emission most likely originates from the
disk surface layer, i.e. the disk photosphere, being directly
illuminated by the central star. To mimic this emission from the disk
surface layer, we used a uniform brightness ring, with an eccentricity
$e$, semi major axis $a$, and orientation $\Phi$, radial width
$\Delta$R, scale-height H$_0$, vertical width $\Delta$H, and surface
brightness I$_\mathrm{ring}$. The ring model is positioned such that
the central emission is in the focus of the ellipse. The central
emission is modeled with a uniform brightness disk with radius
R$_\mathrm{center}$ and surface brightness I$_\mathrm{center}$.
Further, the disk model has a overall inclination $i$ and orientation
$\Theta$. This ``infinite resolution'' mode is then convolved with a 
$\sigma$=0.25 gaussian filter to take into account the limited resolution
of the deconvolved image (0.6''). Our best-fit model (using the 2D fitting procedure
MPFIT2DFUN of IDL package written by \citet{markward}) is shown in
Fig.~\ref{fig-profile} and the resulting fit parameters are listed in
table~\ref{table-1}. We checked also that the overall inverse modeling
is correct by verifying that our solution corresponds also to an minimum
of $\chi^2$ in raw data (non deconvolved) space. 
A sketch of the derived disk structure is shown
in Fig.~\ref{fig-disk}.  The appearance of an emission ring can be
linked to the disk geometry as follows: In a flaring disk model the
amount of radiation intercepted and thus emitted is proportional to
the angle at which the stellar light impinges onto the disk surface.
This angle is a factor of 15 (with the parameters found in the case of
\aba) larger at the ring surface compared to a uniform flaring (wedge)
disk. This implies that a factor of 15 more light is intercepted and
consequently emitted at the disk surface, lifting it above the
detection limit of the CAMIRAS instrument.  This large flaring angle
of the disk surface can also explain the gap in the SW direction, as
the disk surface becomes self shadowed seen under a small inclination
of $-20$\degr.  This kind of ``puffed-up'' structure could be for
instance the result of shadowing instabilities as described in paper
by \citet{2003A&A...398..607D}.  Other possibilities to explain a
sudden increase of brightness at a distance of 260 AU, would be:
\begin{itemize}
\item{} dust sizes segregation \citep{takeuchi2003} but one must infer
a very peculiar structure of the gaseous disk to reproduce the observations.
\item{} PAH photo-dissociation, but it is unlikely to occur to the inferred
 large PAH grains (e.g. \citet{allain1996}).
\item{} an hollow spherical shell and its limb brightening, but
 the off-centering can be hardly explained. 
\end{itemize}  
Remains to explain the origin of the intrinsic eccentricity of the
ring-like structure assumed in the model described above.  Some
massive perturber on an eccentric orbit might produce the detected
gravitational deformations of the ring.  A giant planet is relatively
unlikely to be found one at such a high distance; a not yet detected
brown dwarf could be one possible explanation.

\section{Conclusions}
 We have presented in this paper a thermal infrared image at 20.5
 $\mu$m of the \aba \ dusty disk.  The deconvolved image shows an
 inner resolved structure containing the majority (95 \%) of the
 thermal flux at this wavelength and an unexpected ring-like
 ellipse-shaped structure. This ring is located at an average distance
 of 260 AU from the star, but its most interesting features
 lie in large scale asymmetries and a center that is offset with
 respect to the star.  When modeling the ring structure using a purely
 geometrical model (in which asymmetries are essentially produced by inclination
 and shadowing effects in a flared disk), we end with the conclusion
 that we are in presence of an {\bf intrinsically ellipse-shaped} ring
 that could trace the presence of puffed-up matter from a {\bf flared
 dust disk} containing transiently heated particles.  Self-shadowing
 instabilities perturbating the disk vertical thickness could produce
 such a ring. However, the origin of its intrinsic eccentricity
 remains unclear; massive bodies could be acting in gravitationally
 structuring the disk. Further detailed modelings still need to be
 investigated to fully understand the physics of this intriguing
 structure; further observations, including higher resolution imaging
 in several PAH bands and spatially resolved mid-IR spectroscopy, are
 needed to better understand this object.

\acknowledgements 
We are gratefully indebted to P. Masse, R.Jouan and M.Lortholary for their efficient 
assistance with the CAMIRAS instrument, A. Claret in efficiently supporting us in our
observations, as well as to the staff of
CFHT/Hawaii for their support during the observing runs. 
JB acknowledges financial support by the EC-RTN on ``The Formation
and Evolution of Young Stellar Clusters'' (RTN-1999-00436, HPRN-CT-2000-00155)
The observations preparation was done using the SIMBAD database.

\bibliographystyle{aa}
\bibliography{reference_list_abaur,publication_list_abaur}
 
\end{document}